# Generalized Modal Analysis in Power System with High CIG Penetration: Concept and Quantitative Assessment

Le Zheng, *Member, IEEE*, Jiajie Zheng, and Chongru Liu, *Member, IEEE*

***Abstract*— This paper presents a Generalized Modal Analysis (GMA) concept for the small-signal stability analysis of power systems with high penetration of Converter-Interfaced Generation (CIG). GMA quantitatively assesses interactions between various elements in the power system, offering intuitive and transparent physical interpretations. The method's versatility in selecting physical quantities at different input and output ports makes it broadly applicable. Based on the concept of GMA, the study further defines interaction quantification indices by selecting voltage ports, examining the impact of grid disturbances on power sources and the support from the power sources to the grid at connection points. Numerical simulations on modified 14-bus and 68-bus systems validate GMA's effectiveness in capturing the coupling of the dynamic characteristics between grid elements. This research provides a theoretical foundation and analytical framework for future analyses of power system stability with diverse power sources.**

## I. INTRODUCTION

In recent years, the growing demand for environmentally friendly energy generation has spearheaded the development of large-scale renewable energy generation. As a result, the penetration of Converter-Interfaced Generation (CIG) has significantly increased in power systems. Compared to synchronous generators (SG), CIGs are less robust to disturbances [1]. Often, CIGs induce unstable oscillations with characteristics that markedly differ from those observed in traditional SG-based power systems [2]-[3]. In addition, the CIGs tend to accelerate the propagation of oscillations within the power network [4]. Quantifying the interactions between various apparatus, including CIGs, SGs, and other elements of the grid, has always been a significant challenge in power system small-signal stability analysis.

Modal analysis based on the state-space model (MASS) is a crucial method for evaluating the small-signal stability and dynamic response characteristics of the power system. MASS quantifies the contribution of each state variable to specific modes through participation factors, aiding in identifying the key influencing factors [5]. A technique combining automatic excitation technology with an intrinsic system realization algorithm has been employed in power system modal analysis to estimate modal parameters such as damping ratios and frequencies [6]. Mode selection-based methods have proven effective in calculating coupling modes between synchronous generators (SGs) [7]. The peak selection method in modal analysis identifies resonance frequencies and quality factors from Nyquist plots and underscores the importance of analyzing modal impedance for predicting damping [8]. Despite these advances, existing studies often fall short in determining the most influential components in the system.

On the other hand, impedance model has been widely used in the small-signal stability analysis since the precise analytical state-space model is often unavailable for CIGs [9]-[10]. Impedance model provides some extent of model transparency by resonance peaks and stability margins, for example, several studies have discussed methods for using impedance models to identify key factors influencing system oscillations. The sensitivity of eigenvalues to network component parameters can be used to identify components that affect specific oscillation modes [11]-[12]. Ref. [13] uses loop participation factors and node participation factors to gain deeper insights into the origins of oscillations.

Recently, modal analysis based on the impedance model (MAI) has been developed to trace root causes without requiring detailed internal control information of the power sources. Impedance participation factors are used to identify the power sources that contribute most to system instability. Combined with parameter participation factors, they enable a deeper investigation into the influencing factors of the system [14]-[15]. In addition, the admittance and parameter sensitivity factors are developed to facilitate participation analysis using the entire system's impedance model [16]-[17].

Although MASS and MAI can identify key influencing factors in the system and provide recommendations for system tuning to enhance stability, they lack physical interpretability in the quantitative assessment of the interactions between elements in the network. The main contributions of this paper can be summarized as follows:

(1) A new interpretation of MASS from the transfer function perspective is provided, introducing the concept of Generalized

Manuscript received xx; revised xx. This work was supported by the National Natural Science Foundation of China under Grant 52307095. (*Corresponding author: Chongru Liu*, e-mail: liu.chongru@ncepu.edu.cn).

The authors are with the State Key Lab of Alternate Electrical Power System with Renewable Energy Sources, North China Electric Power University, Beijing 102206, China.

Modal Analysis (GMA). GMA offers an easy way to select any input and output ports of interest and evaluate the coupling relationships between them in relevant mode by calculating the sensitivity of the corresponding eigenvalue to the transfer function from the outputs to the inputs. The method's versatility in selecting physical quantities at different input and output ports makes it broadly applicable.

(2) Based on the concept of GMA, the Voltage Disturbance Margin (VDM) is defined to quantify the impact of the node voltage disturbance to the power sources. The support of the power source to the grid (STG) is further defined and its relationship with other grid strength indicators is discussed. In addition, both model-driven and data-driven methods to compute VDM and STG are proposed.

The structure of this paper is as follows: Section II reinterprets MASS and introduces the concept of GMA. Section III derives VDM and STG using voltage ports with the whole-system impedance matrix. The practical computation method and properties are discussed in Section IV. Section V validates these theories using the 14-bus and 68-bus systems. Section VI concludes the work.

## II. Generalized Modal Analysis

### A. Modal Analysis Based on the State-Space Model

The state-space equation of a linearized power system can be expressed as follows:

$$\begin{cases} \Delta\dot{x} = A\Delta x + B\Delta u \\ \Delta y = C\Delta x + D\Delta u \end{cases} \tag{1}$$

where $\Delta x$ is the state vector, $\Delta u$ is the input vector, and $\Delta y$ is the output vector. The dimensions of $\Delta x$, $\Delta u$ and $\Delta y$ are *n*, *r*, and *m* respectively. *A*, *B*, *C*, and *D* are the state matrix, input matrix, output matrix, and feedforward matrix, respectively.

The full state matrix *A* can be converted to a diagonal one $\Lambda$ by the coordinate transformation $\Delta x = \Phi \Delta z$ .

$$\Lambda = \Psi A \Phi = \mathrm{diag}(\lambda_1 \cdots \lambda_i \cdots \lambda_n) \tag{2}$$

where $\lambda_i$ is the *i*-th eigenvalue of matrix *A*, $\phi_i$ and $\psi_i$ in $\Phi = [\phi_1 \cdots \phi_i \cdots \phi_n]$ and $\Psi = [\psi_1^{\mathrm{T}} \cdots \psi_i^{\mathrm{T}} \cdots \psi_n^{\mathrm{T}}]^{\mathrm{T}}$ are the right and left eigenvector of $\lambda_i$ , respectively. In addition, $\Phi$ and $\Psi$ are inverse matrices of each other, *i.e.*, $\Phi\Psi = \mathbf{I}$, where $\mathbf{I}$ is the unit matrix.

According to [5], the sensitivity of $\lambda_i$ to the *k*-th row and *j*-th column element $a_{kj}$ of the matrix *A* is given by:

$$\frac{\partial \lambda_i}{\partial a_{kj}} = \psi_{ik}\phi_{ji} \tag{3}$$

where $\phi_{ji}$ is the *j*-th element of the right eigenvector $\phi_i$, and $\psi_{ik}$ denotes the *k*-th element of the left eigenvector $\psi_i$ .

The participation matrix $P = [P_1 \cdots P_i \cdots P_n]$ is as follows.

$$P_i = \begin{bmatrix} p_{1i} \\ \vdots \\ p_{ki} \\ \vdots \\ p_{ni} \end{bmatrix} = \begin{bmatrix} \phi_{1i}\psi_{i1} \\ \vdots \\ \phi_{ki}\psi_{ik} \\ \vdots \\ \phi_{ni}\psi_{in} \end{bmatrix} = \begin{bmatrix} \partial\lambda_i / \partial a_{11} \\ \vdots \\ \partial\lambda_i / \partial a_{kk} \\ \vdots \\ \partial\lambda_i / \partial a_{nn} \end{bmatrix} = \mathrm{diag}(\frac{\partial \lambda_i}{\partial A}) \tag{4}$$

where $p_{ki}$ denotes the relative participation of the *k*-th state variable in the *i*-th mode, which equals the sensitivity of $\lambda_i$ to the *k*-th diagonal element $a_{kk}$ of matrix *A*. In other words, modal analysis based on the state space model is equivalent to computing the sensitivity of the eigenvalues to the state matrix. And the sensitivity matrix can be represented by the right and left eigenvectors of the corresponding mode.

$$\frac{\partial \lambda_i}{\partial A} = \begin{bmatrix} \frac{\partial \lambda_i}{\partial a_{11}} & \cdots & \frac{\partial \lambda_i}{\partial a_{1j}} & \cdots & \frac{\partial \lambda_i}{\partial a_{1n}} \\ \vdots & & \vdots & & \vdots \\ \frac{\partial \lambda_i}{\partial a_{i1}} & \cdots & \frac{\partial \lambda_i}{\partial a_{ij}} & \cdots & \frac{\partial \lambda_i}{\partial a_{in}} \\ \vdots & & \vdots & & \vdots \\ \frac{\partial \lambda_i}{\partial a_{n1}} & \cdots & \frac{\partial \lambda_i}{\partial a_{nj}} & \cdots & \frac{\partial \lambda_i}{\partial a_{nn}} \end{bmatrix} = (\phi_i\psi_i)^{\mathrm{T}} \tag{5}$$

### B. Interpretation of MASS from a Transfer Function Perspective

To gain further insights into MASS, we introduce Proposition 1.

*Proposition* 1: For any square matrix *H*(s), let *G*(s) be the its inverse matrix of *H*(s). If $\lambda$ is the solution of $|H(s)| = 0$ , then the sensitivity of *s* to $H(s)$ when $s = \lambda$ is given by

$$\left.\frac{\partial s}{\partial H(s)}\right|_{s=\lambda} = -\mathrm{Res}_{\lambda} G(s)^{\mathrm{T}} \tag{6}$$

in which $\mathrm{Res}_{\lambda} G(s)$ denotes the residue of $G(s)$ at $\lambda$ .

*Proof*: See Appendix A. ■

Let $H(s) = s\mathbf{I} - A$, the determinant of $H(\lambda_i)$ is zero for any eigenvalue $\lambda_i$ of *A* by definition:

$$|H(\lambda_i)| = |\lambda_i \mathbf{I} - A| = 0 \tag{7}$$

Then, (8) can be derived directly from Proposition 1.

$$\frac{\partial \lambda_i}{\partial(\lambda_i \mathbf{I} - A)} = -\left(\mathrm{Res}_{\lambda_i}(s\mathbf{I} - A)^{-1}\right)^{\mathrm{T}} \tag{8}$$

The residue can be further derived as

$$\begin{aligned} \mathrm{Res}_{\lambda_i}(s\mathbf{I} - A)^{-1} &= \lim_{s\to\lambda_i}((s-\lambda_i)\cdot\Phi(s\mathbf{I}-\Lambda)^{-1}\Psi) \\ &= \lim_{s\to\lambda_i}\sum_{j=1}^{n}\phi_j\frac{s-\lambda_i}{s-\lambda_j}\psi_j = \phi_i\psi_i \end{aligned} \tag{9}$$

Combining (5), (8) and (9) leads to

$$\frac{\partial \lambda_i}{\partial A} = (\phi_i\psi_i)^{\mathrm{T}} = \left(\mathrm{Res}_{\lambda_i}(s\mathbf{I} - A)^{-1}\right)^{\mathrm{T}} = -\frac{\partial \lambda_i}{\partial(\lambda_i \mathbf{I} - A)} \tag{10}$$

Apply the Laplace transform to (1)

$$\begin{cases} \Delta x(s) = (s\mathbf{I} - A)^{-1} \left( \Delta x(0) + B\Delta u(s) \right) \\ \Delta y(s) = C\Delta x(s) + D\Delta u(s) \end{cases} \tag{11}$$

where $\Delta x(s)$, $\Delta u(s)$, and $\Delta y(s)$ represent the Laplace transforms of $\Delta x$, $\Delta u$, and $\Delta y$, respectively. $\Delta x(0)$ is the initial state of $\Delta x$. Consider zero-input responses that depend solely on the initial states

$$\Delta x(s) = (s\mathbf{I} - A)^{-1} \Delta x(0) \tag{12}$$

Thus, $H(s) = s\mathbf{I} - A$ is a transfer function from $\Delta x(s)$ to $\Delta x(0)$, illustrating how to trace back to the root cause given current states. Equation (10) signifies that determining the participation factors of states to a mode is equivalent to computing the sensitivity of the corresponding eigenvalue to the transfer function from the current states to the initial ones. This highlights the interpretation of MASS from a transfer function perspective.

### C. Generalized Modal Analysis

MASS focuses on the state variables, and reveals the correlation between the state variables and their initial states. Generally, the coupling relationship between input and output ports of interest can also be approximated by a MASS-like method. For a zero-state response that is independent of the initial state, equation (11) becomes

$$\begin{aligned} \Delta x(s) &= (s\mathbf{I} - A)^{-1} B\Delta u(s) \\ \Delta y(s) &= C\Delta x(s) + D\Delta u(s) \end{aligned} \tag{13}$$

Specifically, we select *m* variables of interest from the state space equation (1) as the input $\Delta x_1$ and another *m* variables from the output vector to form $\Delta y_1$. Then the transfer function of the sub-system can be derived from (13).

$$\Delta y_1(s) = \left( C_1 (s\mathbf{I} - A)^{-1} B_1 + D_1 \right) \Delta x_1(s) = G(s) \Delta x_1(s) \tag{14}$$

where $B_1$ is formed by the corresponding columns from the input matrix $B$, so as $C_1$ and $D_1$. $G(s)$ is a $m \times m$ square matrix. If $G(s)$ is invertible, its inverse matrix $H(s)$ is also a square matrix. The determinant of $H(s)$ is:

$$\left| H(s) \right| = \frac{1}{|G(s)|} = \frac{1}{|C_1 (s\mathbf{I} - A)^{-1} B_1 + D_1|} \tag{15}$$

Substituting $\Phi\Psi = \mathbf{I}$ and $A = \Phi\Lambda\Psi$ into (15):

$$\left| H(s) \right| = \frac{1}{\left| C_1 \Phi (s\mathbf{I} - \Lambda)^{-1} \Psi B_1 + D_1 \right|} \tag{16}$$

where

$$(s\mathbf{I} - \Lambda)^{-1} = \mathrm{diag}\left( \frac{1}{s - \lambda_1} \cdots \frac{1}{s - \lambda_j} \cdots \frac{1}{s - \lambda_n} \right) \tag{17}$$

Therefore, (16) can be simplified to:

$$\begin{aligned} \left| H(s) \right| &= \frac{1}{\left| \sum_{j=1}^{n} \frac{C_1 \phi_j \psi_j B_1}{s - \lambda_j} + D_1 \right|} \\ &= \frac{(s - \lambda_i)^k}{\left| \sum_{j=1}^{n} \frac{(s - \lambda_i) C_1 \phi_j \psi_j B_1}{s - \lambda_j} + (s - \lambda_i) D_1 \right|} \end{aligned} \tag{18}$$

Substitute *s* with an eigenvalue $\lambda_i$, (18) simplifies to:

$$\begin{aligned} \left| H(\lambda_i) \right| &= \frac{(\lambda_i - \lambda_i)^k}{\left| \sum_{j=1}^{n} \frac{(\lambda_i - \lambda_i) C_1 \phi_j \psi_j B_1}{\lambda_i - \lambda_j} + (\lambda_i - \lambda_i) D_1 \right|} \\ &= \frac{(\lambda_i - \lambda_i)^k}{\left| C_1 \phi_j \psi_j B_1 + (\lambda_i - \lambda_i) \left\{ D_1 + \sum_{\substack{j=1 \\ j \neq i}}^{n} \frac{C_1 \phi_j \psi_j B_1}{\lambda_i - \lambda_j} \right\} \right|} \end{aligned} \tag{19}$$

From (19), it is known that $\left| H(s) \right| \neq 0$ holds when $s \neq \lambda_i$, which guarantees $H(s)$ and $G(s)$ are invertible for any $s \neq \lambda_i$. In addition, it has been proved in Appendix B that if the mode is both controllable and observable, the determinant of $H$(s) at $\lambda_i$ is zero, *i.e.*, $\left| H(\lambda_i) \right| = 0$. With $\left| H(\lambda_i) \right| = 0$ satisfaction, we can directly obtain (20) using Proposition 1.

$$\frac{\partial \lambda_i}{\partial H(\lambda_i)} = -\mathrm{Res}_{\lambda_i} G(s)^{\mathrm{T}} \tag{20}$$

Equations (14) and (20) illustrate the fundamental concept of generalized modal analysis (GMA). We can select any input and output ports of interest and evaluate the coupling relationships between them in relevant mode by calculating the sensitivity of the corresponding eigenvalue to the transfer function from the outputs to the inputs. For example, if we inject virtual current disturbances into some apparatus, taking the current injection as the input and the resulting voltage perturbation as the output, we can assess the involvement of the apparatus and network elements by calculating the sensitivity of the mode to the apparatus admittance, which denotes the transfer function from the voltage perturbation to the current injection [16], [15]. In addition, the right term of (20) simplifies the assessment by converting the partial derivative to the computation of the residue of a matrix, providing a more straightforward method to quantitively assess the coupling relationships.

*Remark* 1: Residue itself has significant physical meaning from an interpretative standpoint. The residue of the transfer function at each eigenvalue represents the weighted coefficient of each mode in the output response when the system input is a unit pulse. In particular, the diagonal term in equation (9) is known as the participation factor.

*Remark* 2: The inputs can be further formulated as a linear combination of the state variables, and the same applies to the outputs. The flexibility in selecting ports has broadened GMA's applicability to a much wider range of scenarios.

In the next section, the voltage port of the CIG is utilized to demonstrate GMA in quantitively assessing the support of CIGs to the grid from a small-signal stability standpoint.

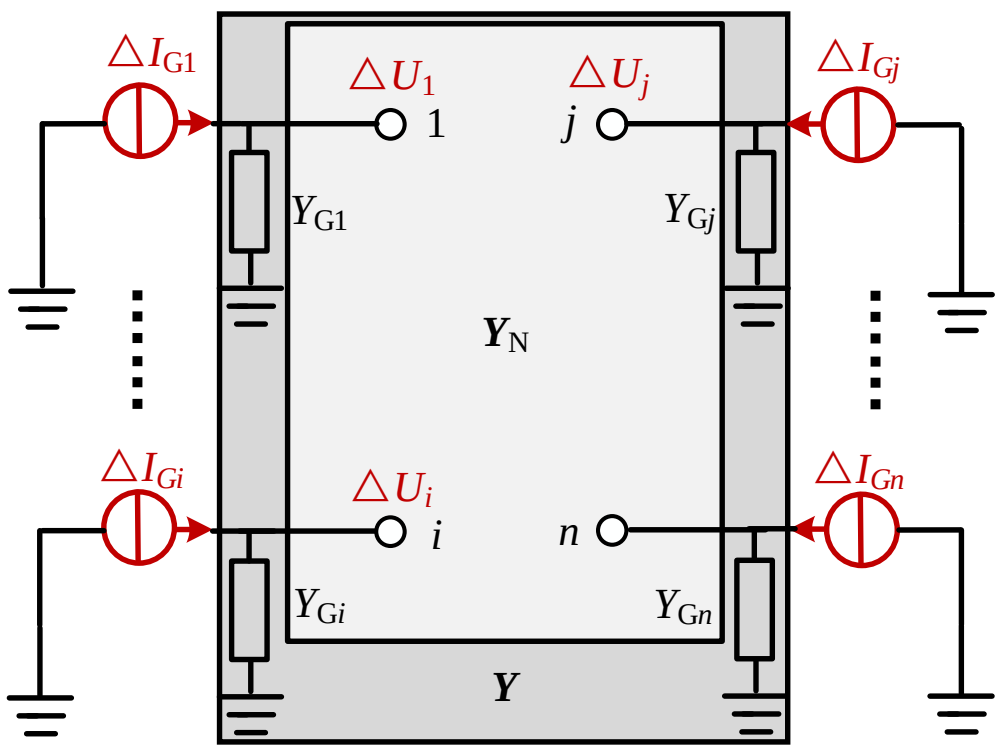

Fig. 1. Illustration of the whole-system impedance model, with virtual current injection and voltage perturbation.

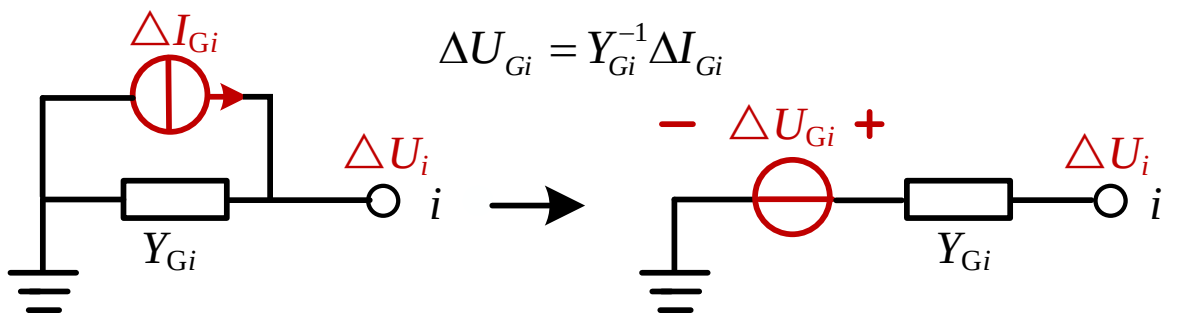

Fig. 2. Norton equivalence and Thevenin equivalence of the power source.

## III. Support of the Power Source to the Grid

### A. Preliminary on Whole-System Impedance Modeling

This part briefly introduces the whole-system impedance modeling method proposed in [9] based on which VDM will be derived. Consider the system with $n$ nodes shown in Fig. 1, $\Delta I_j = \begin{bmatrix} \Delta i_{jd} & \Delta i_{jq} \end{bmatrix}^{\mathrm{T}}$ is the dq component of the injected current at bus $j$ and $\Delta U_j = \begin{bmatrix} \Delta u_{jd} & \Delta u_{jq} \end{bmatrix}^{\mathrm{T}}$ is the voltage perturbation at the same bus. $Y_{Gk}$ represents the admittance model of the power source connected to bus $i$ in dq frame, which is a second-order square matrix:

$$Y_{Gi} = Z_{Gi}^{-1} \tag{21}$$

where $Z_{Gi}$ denotes the impedance model of the power source connected to bus $i$. The impedance or admittance model of a power source can be derived by either from a model-driven method using the state space equations or using a data-driven method with measurements [18]. In particular, $Y_{Gi} = \mathbf{0}_{2\times 2}$ if there is no power source connected to the bus. The power source admittance matrix of the system is (22).

$$\boldsymbol{Y}_G = \mathrm{diag}(Y_{G1} \cdots Y_{Gi} \cdots Y_{Gn}) \in \mathbb{R}^{2n\times 2n} \tag{22}$$

The node admittance matrix representing the network connection and dynamics in $s$ domain is

$$\boldsymbol{Y}_N = \begin{bmatrix} Y_{N11} & \cdots & Y_{N1j} & \cdots & Y_{N1n} \\ \vdots & & \vdots & & \vdots \\ Y_{Ni1} & \cdots & Y_{Nij} & \cdots & Y_{Nin} \\ \vdots & & \vdots & & \vdots \\ Y_{Nn1} & \cdots & Y_{Nnj} & \cdots & Y_{Nnn} \end{bmatrix} \tag{23}$$

where $Y_{Nij} \in \mathbb{R}^{2\times 2}$ denotes the mutual admittance between buses $i$ and $j$, and $Y_{Nii} \in \mathbb{R}^{2\times 2}$ denotes the self-admittance of bus $i$.

Take the injected current disturbance $\Delta I_i$ at each bus as input, and the corresponding voltage perturbation $\Delta U_i$ as output:

$$\begin{cases} \Delta \boldsymbol{I} = [\Delta I_1 \cdots \Delta I_i \cdots \Delta I_n]^{\mathrm{T}} \\ \Delta \boldsymbol{U} = [\Delta U_1 \cdots \Delta U_i \cdots \Delta U_n]^{\mathrm{T}} \end{cases} \tag{24}$$

The closed-loop transfer function matrix from input to output can be derived [9].

$$\Delta \boldsymbol{U} = (\boldsymbol{I} + \boldsymbol{Z}_N \boldsymbol{Y}_G)^{-1} \boldsymbol{Z}_N \Delta \boldsymbol{I} = \boldsymbol{Z} \Delta \boldsymbol{I} \tag{25}$$

$$\boldsymbol{Z} = (\boldsymbol{I} + \boldsymbol{Z}_N \boldsymbol{Y}_G)^{-1} \boldsymbol{Z}_N = (\boldsymbol{Y}_G + \boldsymbol{Y}_N)^{-1} = \boldsymbol{Y}^{-1} \tag{26}$$

where $\boldsymbol{Z}_N = \boldsymbol{Y}_N^{-1}$ is the node impedance matrix of the network. $\boldsymbol{Z}$ and $\boldsymbol{Y}$ are called the whole-system impedance matrix and the whole-system admittance matrix, respectively. Thus, the dynamics of both the power source and the network are included in the whole-system impedance and admittance matrices.

### B. Voltage Disturbance Margin

The voltage perturbation of the power source $\Delta U_{Gi}$ can be derived using the Thevenin-Norton theorem, as shown in Fig. 2 and (27).

$$\Delta U_{Gi} = Y_{Gi}^{-1} \Delta I_{Gi} \tag{27}$$

If bus $i$ is not connected to an actual power source, let $Y_{Gi} = \mathrm{diag}(\varepsilon, \varepsilon)$, where $\varepsilon$ is a small number, ensuring the invertibility of $Y_{Gi}$. Rewrite (27) in matrix form:

$$\Delta \boldsymbol{U}_G = \boldsymbol{Y}_G^{-1} \Delta \boldsymbol{I}_G \tag{28}$$

where $\Delta \boldsymbol{U}_G$ and $\Delta \boldsymbol{I}_G$ denote the voltage and current vector of all the power sources, respectively. $\Delta \boldsymbol{I}_G$ is the injected current disturbance, *i.e.*, $\Delta \boldsymbol{I}_G = \Delta \boldsymbol{I}$ and $\Delta \boldsymbol{I}_G = \boldsymbol{Y} \Delta \boldsymbol{U}$. Substituting $\Delta \boldsymbol{I}$ into (28) leads to

$$\Delta \boldsymbol{U}_G = \boldsymbol{Y}_G^{-1} \boldsymbol{Y} \Delta \boldsymbol{U} = (\boldsymbol{Z} \boldsymbol{Y}_G)^{-1} \Delta \boldsymbol{U} \tag{29}$$

Equation (29) describes the influence of the network voltage perturbation to that of the power source with transfer function $\boldsymbol{Y}_G^{-1} \boldsymbol{Y}$. For any eigenvalue $\lambda_k$ of the system, the determinant of the whole-system admittance matrix at $\lambda_k$ is zero [16]. Then the determinant of the transfer function is also zero.

$$|\boldsymbol{Y}_G^{-1}(\lambda_k) \cdot \boldsymbol{Y}(\lambda_k)| = |\boldsymbol{Y}_G^{-1}(\lambda_k)| \cdot |\boldsymbol{Y}(\lambda_k)| = 0 \tag{30}$$

Using Proposition 1, the sensitivity of the eigenvalue to the transfer function can be computed using the residue of the inverse matrix, as shown in (31). It reveals the interactions between the network and power sources in different oscillatory modes through voltage ports.

$$\begin{aligned} \frac{\partial \lambda_k}{\partial [\boldsymbol{Y}_G^{-1}(\lambda_k) \cdot \boldsymbol{Y}(\lambda_k)]} &= -\mathrm{Res}_{\lambda_k} [\boldsymbol{Z}(s) \cdot \boldsymbol{Y}_G(s)]^{\mathrm{T}} \\ &= -\boldsymbol{Y}_G^{\mathrm{T}}(\lambda_k) \cdot \mathrm{Res}_{\lambda_k} \boldsymbol{Z}^{\mathrm{T}}(s) \end{aligned} \tag{31}$$

To delve deeper into the interactions between specific ports, we can expand (31) to obtain (32), which denotes the modal voltage sensitivity between node $j$ and power source $i$.

$$\frac{\partial \lambda_k}{\partial [\boldsymbol{Y}_G^{-1}(\lambda_k)\cdot \boldsymbol{Y}(\lambda_k)]_{ij}} = -Y_{Gi}^{\mathrm{T}}(\lambda_k)\cdot \mathrm{Res}_{\lambda_k} \boldsymbol{Z}_{ji}^{\mathrm{T}}(s) \tag{32}$$

where $\boldsymbol{Z}_{ji}(s) = \begin{bmatrix} Z_{\mathrm{dd}ji} & Z_{\mathrm{dq}ji} \\ Z_{\mathrm{qd}ji} & Z_{\mathrm{qq}ji} \end{bmatrix}$ is the mutual impedance between node $j$ and $i$ extracted from the whole-system impedance matrix $\boldsymbol{Z}$.

It is noteworthy that if bus $i$ is not connected to an actual power source, then (32) is zero. The sensitivity of $\lambda_i$ to $(\boldsymbol{Y}_G^{-1}\boldsymbol{Y})_{ij}$ is zero, thus the assumption in (27) does not affect the calculation. $[\boldsymbol{Y}_G^{-1}\boldsymbol{Y}]_{ij}$ is the transfer function from the voltage perturbation of bus $j$ to that of the power source $i$. Since (32) is a two-dimensional matrix, the Frobenius norm is used to approximate the magnitude of the sensitivity as:

$$\left| \frac{\partial \lambda_k}{\partial [\boldsymbol{Y}_G^{-1}(\lambda_k)\cdot \boldsymbol{Y}(\lambda_k)]_{ij}} \right|_F = \left| \mathrm{Y}_{Gi}^{\mathrm{T}}(\lambda_k)\cdot \mathrm{Res}_{\lambda_k} \boldsymbol{Z}_{ji}^{\mathrm{T}}(s) \right|_F \tag{33}$$

where $|*|_F$ means applying the Frobenius norm to the matrix. By using an infinitesimal to replace the partial derivative in the above equation, we obtain

$$\left| \Delta [\boldsymbol{Y}_G^{-1}(\lambda_k)\cdot \boldsymbol{Y}(\lambda_k)]_{ij} \right|_F = \frac{|\Delta \lambda_k|_F}{\left| Y_{Gi}^{\mathrm{T}}(\lambda_k)\cdot \mathrm{Res}_{\lambda_k} \boldsymbol{Z}_{ji}^{\mathrm{T}}(s) \right|_F} \tag{34}$$

The left term of (34) measures the extent to which the power source $i$ can tolerate disturbances from node $j$. Therefore, the Voltage Disturbance Margin (VDM) is defined as the maximum estimation of (34).

$$\begin{aligned} \mathrm{VDM}_{ij}(\lambda_k) &= \max \left\{ \left| \Delta [\boldsymbol{Y}_G^{-1}(\lambda_k)\cdot \boldsymbol{Y}(\lambda_k)]_{ij} \right|_F \right\} \\ &= \frac{-\sigma_k}{\left| Y_{Gi}^{\mathrm{T}}(\lambda_k)\cdot \mathrm{Res}_{\lambda_k} \boldsymbol{Z}_{ji}^{\mathrm{T}}(s) \right|_F} \end{aligned} \tag{35}$$

where $\sigma_k$ is the real part of $\lambda_k = \sigma_k + \mathrm{j}\omega_k$ . Thus, $-\sigma_k$ represents the maximum possible change in $|\Delta \lambda_k|_F$ .

### C. Support of the Power Source to the Grid

Recall that the diagonal elements of the sensitivity of the eigenvalue to the state matrix denotes the participation factors. Therefore, we further define the support of the power source to the grid (STG) by assessing the diagonal elements $\mathrm{VDM}_{ii}$ at each individual power source $i$ as (36).

$$\mathrm{STG}_i = \min_{\lambda_k \in \lambda}\{\mathrm{VDM}_{ii}\} = \min_{\lambda_k \in \lambda}\{\frac{-\sigma_k}{\left| Y_{Gi}^{\mathrm{T}}(\lambda_k)\cdot \mathrm{Res}_{\lambda_k} \boldsymbol{Z}_{ii}^{\mathrm{T}}(s) \right|_F}\} \tag{36}$$

where $\min_{\lambda_k \in \lambda}\{*\}$ denotes the minimum value calculated among a subset of modes of particular interest. This is because if a power source can observe multiple modes, it is highly sensitive to oscillations at different frequencies. Thus, STG is determined by the minimum VDM value calculated in different modes.

In general, it is not necessary to account for every eigenvalue. The selection of the subset of modes follows two criteria. First, only eigenvalues with small real parts close to the imaginary axis are considered, as these lightly damped modes are most likely to be excited. These modes usually appear as peaks in the spectrum of the system node impedance matrix. Second, oscillations with high-frequency do not easily propagate through the network, which further narrows the selection.

Both VDM and STG are local small-signal indicators representing different ports in multi-infeed power systems. These indicators inherently reflect the interactions between power sources through the grid. Additionally, VDM and STG are computed using the power source admittance matrix and the node impedance matrix. As a result, they naturally encompass the dynamics of the CIG control loops and the power networks.

The sign of VDM and STG indicates whether the system is small-signal stable, and the magnitude indicates the stability margin from different ports. Specifically, the value of $\mathrm{VDM}_{ij}$ indicates the extent to which power source $i$ can tolerate disturbances from node $j$ in the mode, whereas the value of $\mathrm{STG}_i$ indicates the support provided by power source $i$ to the grid to enhance small-signal stability. Larger values of VDM and STG suggest that the power source is less sensitive to disturbances and can provide stronger support to the grid, respectively.

## IV. Properties of VDM and STG

### A. Practical Method to Compute VDM and STG

Since STG is derived from VDM, only the computation method of VDM will be discussed below. Calculating $\mathrm{VDM}_{ij}(\lambda_k)$ requires the values of $\sigma_k$, $\mathrm{Res}_{\lambda_k} \boldsymbol{Z}_{ji}(s)$, and $Y_{Gi}(\lambda_k)$. The first two can be obtained from the elements of the whole-system node impedance matrix $\boldsymbol{Z}_{ji}(\lambda_k)$. $\boldsymbol{Z}_{ji}(\lambda_k)$ and $Y_{Gi}(\lambda_k)$ can be obtained through two methods.

*1) Model-driven method*: When precise analytical models of the dynamic behavior of the CIGs are available, power source admittance matrix of the system $\boldsymbol{Y}_G$ can be calculated through linearization. Combined with the network node admittance matrix $\boldsymbol{Y}_N$, the whole-system impedance matrix $\boldsymbol{Z}$ can be obtained using (26). This yields the necessary components for $\mathrm{VDM}_{ij}(\lambda_k)$, which is the method used in the case study of this paper.

*2) Data-driven method*: In some cases, there is no detailed admittance model available, only a set of frequency response data for the power source, which is called the black-box characteristics. In this case, vector fitting techniques can be used to obtain $Y_{Gi}(\lambda_k)$ from $Y_{Gi}(\mathrm{j}\omega)$ [19]. $\boldsymbol{Z}_{ji}(\mathrm{j}\omega)$ can be obtained by injecting a current disturbance at node $i$ and measuring the voltage frequency response at node $j$. $\mathrm{Res}_{\lambda_k} \boldsymbol{Z}_{ji}(s)$ and $\sigma_k$ can be derived from the vector fitting results of $\boldsymbol{Z}_{ji}(\mathrm{j}\omega)$ . This method enables real-time online calculation of VDM and STG. Additionally, the eigensystem realization algorithm (ERA) provides a means to directly obtain the full system node admittance matrix Y from time-domain data [20], allowing for the further derivation of VDM.

It is noteworthy that the small-signal model of the system depends on the operating point; thus, it is recommended to estimate VDM and STG at various reliable operating points or periodically in real-time.

### B. Connections to Other Grid Strength Indices

The short circuit ratio (SCR) has traditionally been employed to evaluate grid strength, with numerous studies utilizing it to analyze the small-signal stability issues introduced by CIGs [21]-[22]. Given this context, it is crucial to elucidate the connection between STG, a newly introduced indicator of small-signal stability, and the commonly used SCR.

As illustrated in Fig. 3, $Z_g$ and $Z_s$ represent the equivalent impedances of the grid and the CIG, respectively, whereas $U_g$ and $U_i$ denote the grid voltage and the voltage at bus $i$, respectively. Typically, both $U_g$ and $U_i$ are 1 p.u..

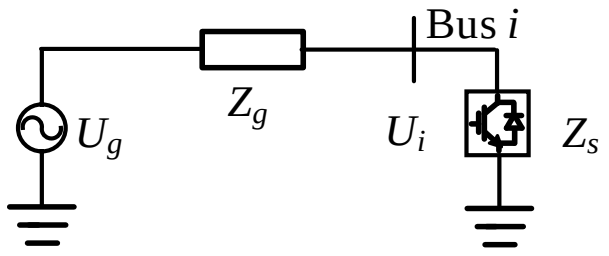

Fig. 3. A CIG connected to a single power source infinite bus system.

According to the definition of SCR:

$$\mathrm{SCR} = \frac{S_{\mathrm{ac}}}{P_{Ni}} \tag{37}$$

where $S_{\mathrm{ac}}$ and $P_{Ni}$ are the short-circuit capacity of bus $i$ and the rated active power of the CIG, respectively.

According to the grid code, the reactive power of the CIG can be ignored under normal operating conditions. Therefore, $S_{\mathrm{ac}}$ and $P_{Ni}$ are calculated as follows:

$$S_{\mathrm{ac}} = \frac{U_g^2}{|Z_g|} = \frac{1}{|Z_g|}, P_{Ni} = \frac{U_i^2}{|Z_s|} = \frac{1}{|Z_s|} \tag{38}$$

Substituting (38) into (37) gives:

$$\mathrm{SCR} = \frac{|Z_s|}{|Z_g|} \tag{39}$$

From (34), (35), and (36), it can be seen that:

$$\mathrm{STG}_i = \min_{\lambda}\left\{\max\left\{\left|\Delta[\boldsymbol{Y}_G^{-1}\boldsymbol{Y}]_{ii}\right|_F\right\}\right\} \tag{40}$$

SCR only considers steady-state components at the base frequency, substituting (39) into (40):

$$\begin{aligned} \mathrm{STG}_{i\,\lambda=\mathrm{j}\omega_{base}} &= \max\Delta\left\{|Z_s|\cdot\left(\frac{1}{|Z_s|}+\frac{1}{|Z_g|}\right)\right\} \\ &= \max\Delta\{1+\mathrm{SCR}\} \\ &= \max\{\Delta\mathrm{SCR}\} \end{aligned} \tag{41}$$

where $\omega_{base}$ denotes the base angular frequency. It is clear that the maximum variation of SCR represents a special case of $\mathrm{STG}_i$ evaluated solely at the base frequency. Thus, STG provides significantly more information than SCR.

In addition, the site-dependent short circuit ratio (SDSCR) [23], the voltage stiffness (VS) [24], and the generalized short circuit ratio (GSCR) [25]-[26] are also used to indicate grid strength and evaluate small-signal stability. SDSCR is a natural extension of SCR, and VS treats converter impedance as open or short circuits depending on the control mode of the power source. They can only implicitly reflect the impacts of the control loops of CIGs. Regarding GSCR, it uses the system's minimum eigenvalue as a global indicator to measure grid strength, which results in a loss of detailed information at different locations of the grid.

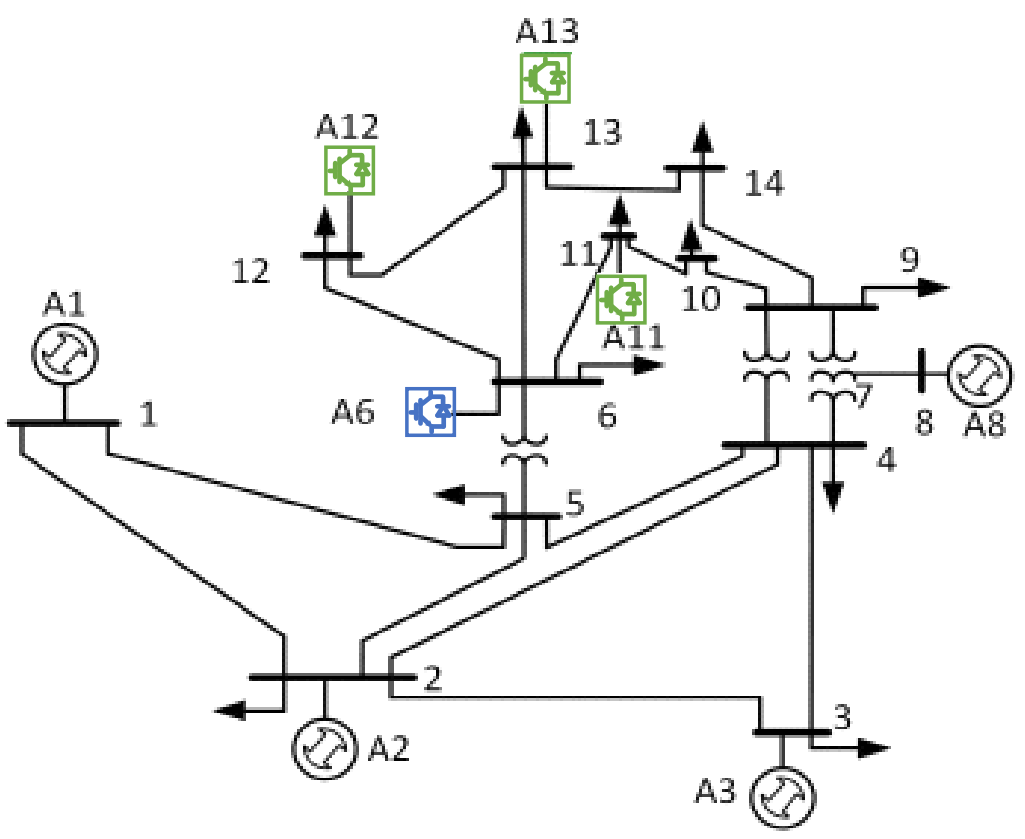

Fig. 4. The modified 14-bus system.

TABLE I
LINE IMPEDANCE AND VDM

| | Line 6-11 | Line 6-12 | Line 6-13 |
|---|---|---|---|
| Line Impedance (p.u.) | 0.095+j0.199 | 0.123+j0.256 | 0.066+j0.130 |
| VDM | 0.365 | 0.249 | 0.291 |

## V. CASE STUDIES

This section validates the performance of VDM and STG through two case studies. First, the modified 14-bus system is used to demonstrate that VDM can characterize the impact of the grid on power sources and effectively capture the interaction between power sources through the grid. Secondly, the modified 68-bus system is utilized to illustrate the role of STG in evaluating the support of power sources to enhance small-signal stability in large-scale power systems.

### A. 14-bus System

Fig. 4 shows the modified 14-bus system. SGs are connected to buses 1, 2, 3, and 8. A CIG operating in grid-forming mode (GFM) is connected to bus 6, whereas three CIGs operating in grid-following mode (GFL) are connected to buses 11, 12, and 13. The GFL units, labeled A11, A12, and A13 have identical control parameters except for the current loop bandwidth, which are set to 400 Hz, 300 Hz, and 350 Hz, respectively. Table I presents the impedance of the lines connecting the GFM to the GFLs, with the line 6-12 having the highest impedance. If the dynamics of the power sources are not considered, the impact on A6 should follow the order of bus 13 > bus 11 > bus 12, based on the principle that a shorter electrical distance typically implies a greater impact on system dynamics.

There is a mode located in the sub-synchronous frequency band with an eigenvalue of $\lambda_k = -4.09 + \mathrm{j}30.06\mathrm{Hz}$, which can also be validated by the Bode diagrams of $\boldsymbol{Z}_{ij}$ in the d-d axis, as shown in Fig. 5(a). The spectral peak of $Z_{\mathrm{dd}12\text{-}6}$ having the largest magnitude, indicates that bus 12 affects A6 the most for this mode. We then evaluate the impact of the voltage disturbances from the three GFL units on the GFM unit A6 using VDM, as shown in Table I. $\mathrm{VDM}_{6\text{-}12}$ is the smallest among the three VDMs, suggesting that the dynamics of bus 12 has the greatest impact on A6. This is consistent with the time-domain simulation results shown in Fig. 5(d), where the oscillation peaks are the largest when the active load of bus 12 increases by 1 p.u. at 0.1s. Thus, both VDM and simulation results confirm that the impact on A6 follows the order of bus 12 > bus 13 > bus 11. This is contrary to the conclusions from the electrical distance analysis, which reflects the system impedance near base frequency and treats the power source either open circuit or short circuit based on the synchronization strategy [24], [27]. Therefore, it is crucial to consider the effect of power sources when analyzing the dynamics of power systems with high CIG penetration, especially for the modes away from the base frequency.

TABLE II
THE VDM$_{ij}$ VALUES UNDER DIFFERENT A6 VOLTAGE LOOP BANDWIDTHS

| Voltage Loop Bandwidth (Hz) | VDM$_{6-11}$ | VDM$_{6-12}$ | VDM$_{6-13}$ |
|---|---|---|---|
| 240 | 0.259 | 0.208 | 0.221 |
| 360 | 0.479 | 0.364 | 0.394 |

TABLE III
STG IN 68-BUS SYSTEM BEFORE AND AFTER PARAMETER ADJUSTMENTS

| | Before | After | | Before | After |
|---|---|---|---|---|---|
| STG$_{15}$ | 0.281 | 0.466 | STG$_{39}$ | 0.255 | 0.255 |
| STG$_{17}$ | 4.308 | 4.301 | STG$_{43}$ | 0.711 | 0.711 |
| STG$_{26}$ | 3.411 | 4.456 | STG$_{58}$ | 2.902 | 3.454 |
| STG$_{28}$ | 1.036 | 1.702 | STG$_{59}$ | 2.081 | 2.421 |
| STG$_{29}$ | 0.155 | 1.129 | STG$_{60}$ | 0.663 | 0.774 |

To further analyze the impact of the parameters, the voltage loop bandwidth of A6 was adjusted to 240 Hz and 360 Hz, respectively. The spectra of the corresponding $Z_{dd}$ are shown in Fig. 5 (b) and (c), and the VDM is shown in Table II. A decrease in the voltage loop bandwidth results in an increase in the spectral peak and a decrease in VDM, indicating a larger impact on A6. The simulations in Fig. 5 (e) and (f) validate the effectiveness of VDM in quantitively assessing the modal interaction among different elements.

## *B. 68-bus System*

Fig. 6 presents the modified 68-bus system, where the GFM units are highlighted in blue, and the GFL units in green. Notably, the droop gain of GFM15 is intentionally set to a higher value of 0.01, with its voltage loop bandwidth reduced to 300 Hz. Similarly, the current loop bandwidth of GFL29 is intentionally reduced to 200 Hz.

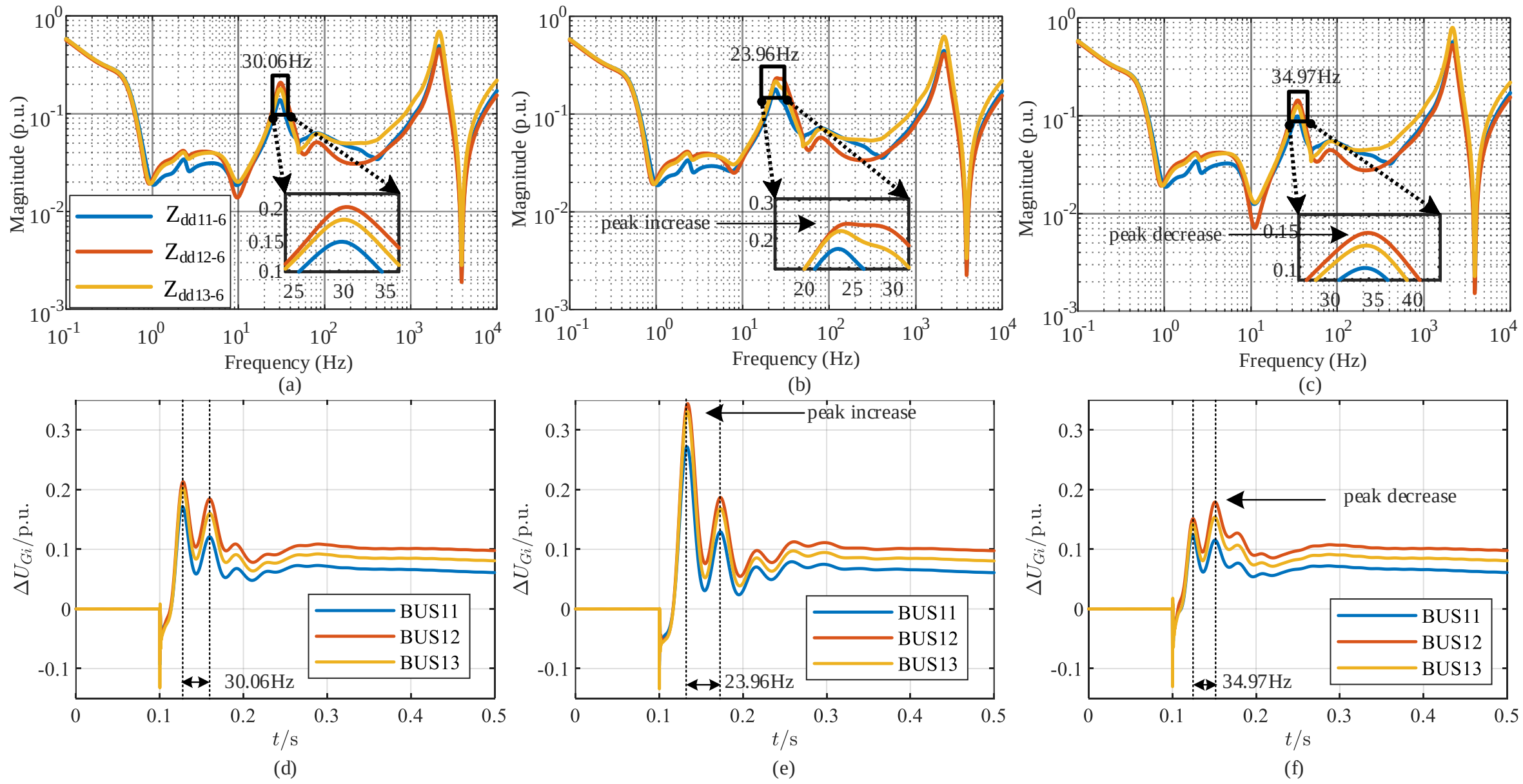

Fig. 5 (a)(b)(c) Bode plots of $Z_{dd11-6}$, $Z_{dd12-6}$, and $Z_{dd13-6}$ at 300 Hz, 240 Hz, and 360 Hz for the A6 voltage loop, respectively. (d)(e)(f) The changes in the voltage magnitude of power source A6 at 300 Hz, 240 Hz, and 360 Hz when the active load at buses 11, 12, and 13 increases by 1 p.u. at t=0.1 s.

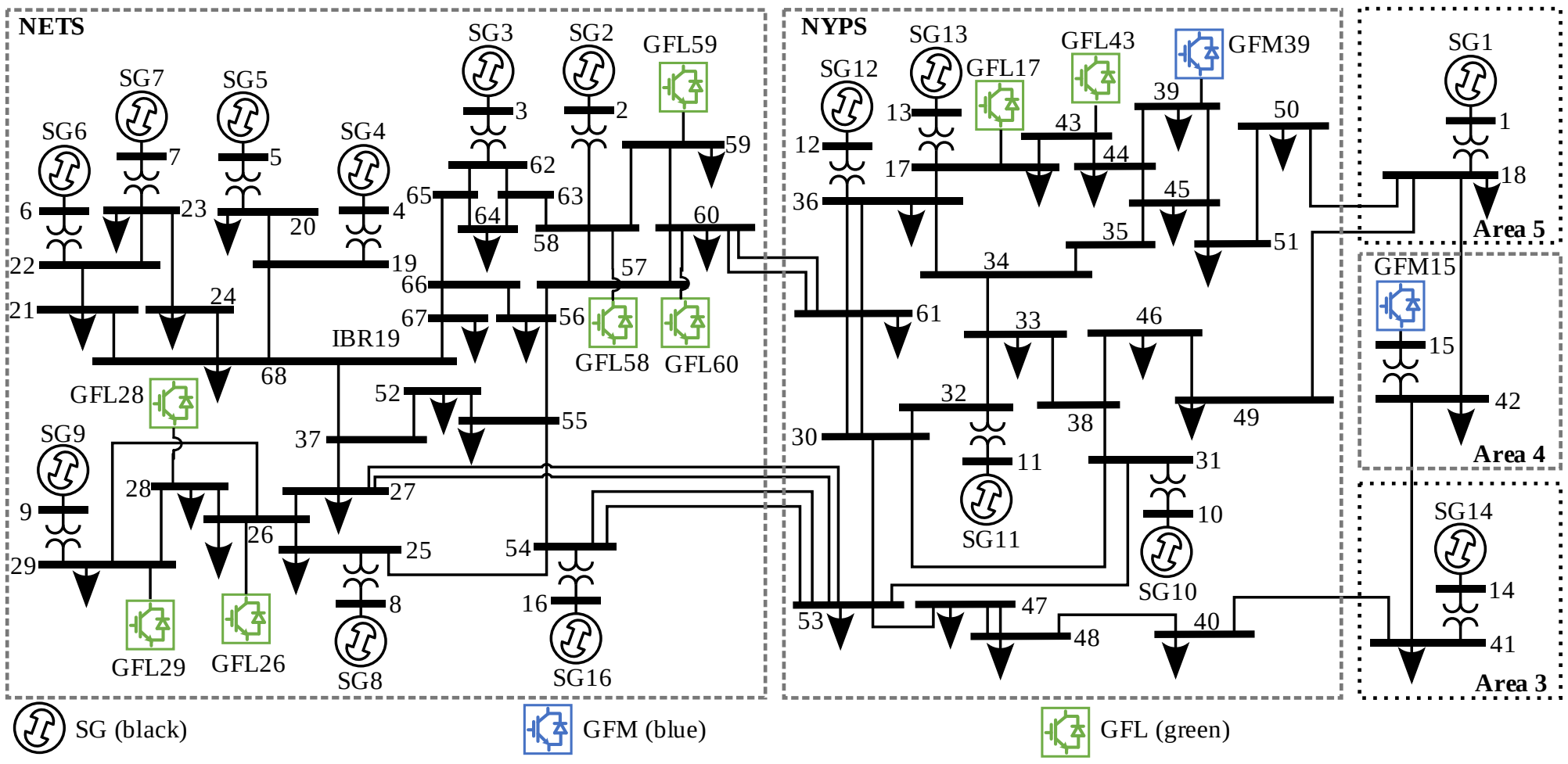

Fig. 6 The modified 68-bus system.

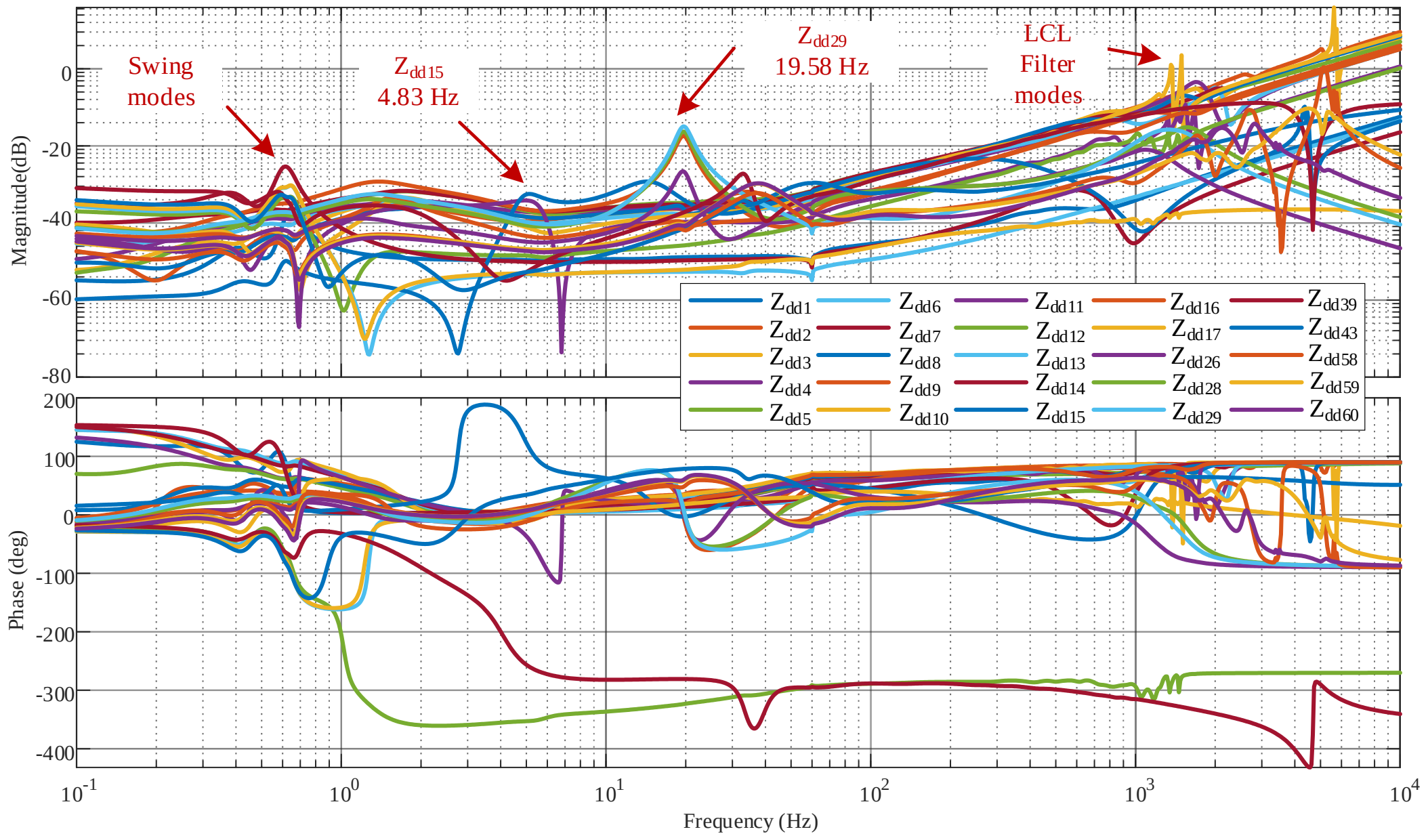

Fig. 7 The Bode plots of the dd elements of the impedance matrix for the entire system corresponding to the power sources.

Fig. 7 displays the spectra of the diagonal elements of the whole-system impedance matrix for the power sources. It can be observed that $Z_{dd15}$ has a peak at 4.83 Hz and $Z_{dd29}$ exhibits a significant peak at 19.58 Hz, indicating that GFM15 and GFL29 are susceptible to oscillations at their respective frequencies. The STG values of all the CIG nodes are displayed in Table III. Specifically, the values of $STG_{15}$ and $STG_{29}$ are 0.281 and 0.155, which are relatively low compared to other CIGs. This indicates that the support from GFM15 and GFL29 to the grid is low, making the system prone to oscillation from the small-signal stability point of view.

The parameters of the two CIGs were then readjusted as follows: the droop gain and voltage loop bandwidth of GFM15 were set to 0.005 and 450 Hz, respectively, and the current loop bandwidth of GFL29 was set to 400 Hz. After these adjustments, the values of $STG_{15}$ and $STG_{29}$ increased significantly to 0.466 and 1.129, respectively, indicating that GFM15 and GFL29 can now provide stronger disturbance resilience support to the system. Fig. 8 (a) and (b) shows the voltage amplitude changes of GFM15 and GFL29 when the active load at buses 15 and 29 increases by 1 p.u., before and after parameter adjustments. The simulation demonstrates the effectiveness of STG in describing the support of power sources in enhancing system stability. Based on extensive simulations of the 68-bus system, it is suggested to use STG values of 0.4 and 1 as the boundary values for distinguishing between strong and weak support of GFM and GFL units, respectively.

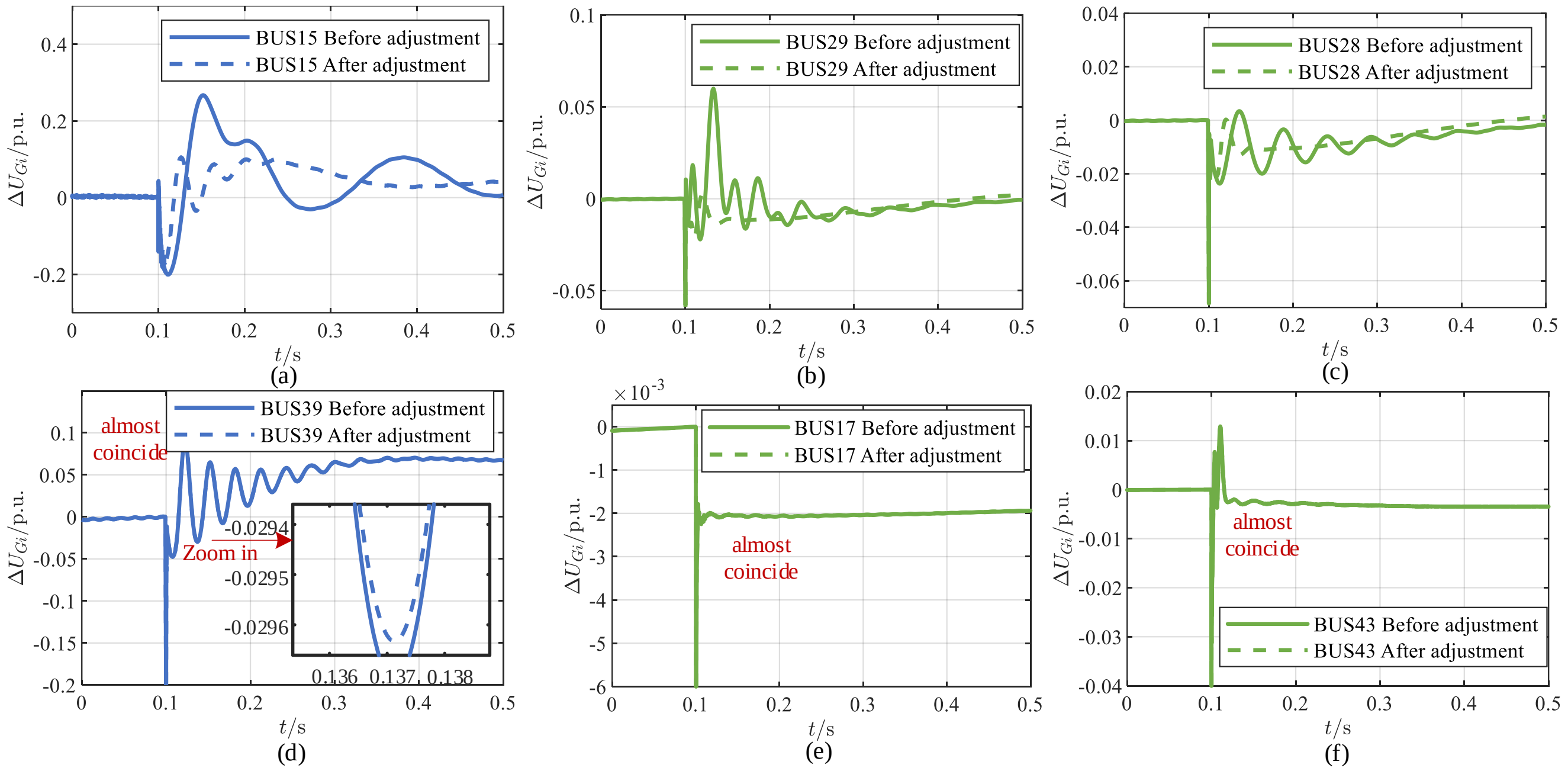

Fig. 8 The change in converter voltage at the bus connection point due to an increase of 1 p.u. in active power load at 0.1 seconds. (a) Bus15. (b) Bus29. (c) Bus28. (d) Bus39. (e) Bus17. (f) Bus43.

Furthermore, although only the control parameters of GFM15 and GFL29 were adjusted, the value of STG28 changed from 1.036 to 1.702. Setting the same fault at bus 28, Figure 8(c) indicates that the oscillation of GFL28 quickly subsided after adjusting the parameters of GFL29. Therefore, STG can fully reflect the mutual influence of converter dynamics. It is worth noting that the values of STG39, STG17, and STG43 remained almost unchanged, indicating a weak mutual influence between the converters in the NYPS region and GFM15 and GFL29 as shown in Figure 6. Setting the same fault at buses 17, 43, and 39 respectively, the simulation results are shown in Figures 8(d), (e), and (f). It can be observed that the waveform of GFM39 exhibits significant oscillations, while the waveform of GFL17 shows minimal fluctuations. Additionally, the waveforms of GFM39, GFL17, and GFL43 are almost coincident before and after the parameter adjustments. The simulation results clearly demonstrate the effectiveness of the indicators.

## VI. Conclusion

This paper first interprets modal analysis based on the state-space model (MASS) and participation factors from a transfer function perspective. Through rigorous mathematical derivation, the concept of Generalized Modal Analysis (GMA) is proposed, offering an intuitive way to quantitatively assess the interactions of various elements in power systems by using the sensitivity of any mode of interest to the transfer function. Specifically, MASS is proven to be a special case of GMA.

The Voltage Disturbance Margin (VDM) and Support To the Grid (STG) are proposed as new metrics to indicate the impact and support of power sources to the grid from a small-signal stability standpoint, based on the concept of GMA by choosing the voltage disturbance of various nodes as the input and output ports. Compared to the commonly used Short Circuit Ratio (SCR), VDM and STG can provide modal analysis between any apparatus at any specific mode of interest. Additionally, the dynamics of CIG control loops are explicitly and inherently considered in the derivation. Case studies have verified the effectiveness of VDM and STG in assessing the oscillatory risk under various operating conditions.

With a solid theoretical foundation and practical calculation methods, GMA possesses unique advantages in quantitatively assessing the interactions of elements in power systems with high CIG penetration. The method's versatility in selecting physical quantities at different input and output ports makes it broadly applicable. Future work will explore more practical applications of GMA in CIG planning, operation, and control.

## Appendix A

### Proof of Proposition 1

The determinant $|H(s)|$ of any square matrix $H(s)$ satisfies:

$$|H(s)| = \sum_{i=1}^{n} H_{ij} E_{ij} \tag{A1}$$

where $E_{ij}$ is the algebraic cofactor of the element $H_{ij}$ in the $i$-th row and $j$-th column of $H(s)$. From (A1), we obtain:

$$\frac{\partial |H(s)|}{\partial H_{ij}} = E_{ij} \tag{A2}$$

If $\lambda$ is the solution to $|H(s)| = 0$, and $H_{ij}$ is perturbed by a value $\Delta H_{ij}$, then we have:

$$|H(\lambda + \Delta\lambda \ , \ H_{ij} + \Delta H_{ij})| = 0 \tag{A3}$$

where $\Delta\lambda$ is the change in the solution.

Expanding (A3) using Taylor series and neglecting higher-order infinitesimals:

$$|H(\lambda)| + \frac{\partial |H(\lambda)|}{\partial \lambda} \Delta\lambda + \frac{\partial |H(\lambda)|}{\partial H_{ij}} \Delta H_{ij} = 0 \tag{A4}$$

Substituting (A2) and (A3) into (A4):

$$\Delta\lambda = -\left( \frac{\partial |H(\lambda)|}{\partial \lambda} \right)^{-1} E_{ij} \Delta H_{ij} \tag{A5}$$

$$\frac{\partial \lambda}{\partial H_{ij}} = \frac{\Delta\lambda}{\Delta H_{ij}} = -\left( \frac{\partial |H(\lambda)|}{\partial \lambda} \right)^{-1} E_{ij} \tag{A6}$$

$$\begin{aligned} \frac{\partial \lambda}{\partial H(\lambda)} &= -\left( \frac{\partial |H(\lambda)|}{\partial \lambda} \right)^{-1} \begin{bmatrix} E_{11} & \cdots & E_{1j} & \cdots & E_{1n} \\ \vdots & & \vdots & & \vdots \\ E_{i1} & \cdots & E_{ij} & \cdots & E_{in} \\ \vdots & & \vdots & & \vdots \\ E_{n1} & \cdots & E_{nj} & \cdots & E_{nn} \end{bmatrix} \\ &= -\left( \frac{\partial |H(\lambda)|}{\partial \lambda} \right)^{-1} H(\lambda)^* \end{aligned} \tag{A7}$$

where $\partial\lambda / \partial H$ is the sensitivity matrix of $\lambda$ with respect to $H(\lambda)$, and * denotes the adjoint matrix.

Let matrix $G(s)$ be the inverse matrix of $H(s)$, and the residue of $G(s)$ at $\lambda$ is:

$$\mathrm{Res}_{\lambda} G(s) = \lim_{S \to \lambda} ((s-\lambda) \frac{H(s)^*}{|H(s)|}) = \left( \frac{\partial |H(\lambda)|}{\partial \lambda} \right)^{-1} H(\lambda)^* \tag{A8}$$

Combining (52) and (53), we obtain:

$$\left. \frac{\partial s}{\partial H(s)} \right|_{s=\lambda} = -\mathrm{Res}_{\lambda} G(s)^{\mathrm{T}} \tag{A9}$$

Therefore, the sensitivity of $\lambda$ with respect to $H(\lambda)$ equals the negative of the residue of $G(s)^{\mathrm{T}}$ at $\lambda$. ■

## Appendix B

When the $i$-th mode is controllable and observable, we have $\psi_i B_1 \neq 0$ and $C_1 \phi_i \neq 0$, thus (18) simplifies to:

$$|H(\lambda_i)| = \frac{(\lambda_i - \lambda_i)^m}{|C_1 \phi_i \psi_i B_1|} \tag{B1}$$

At this point, we have:

$$\mathrm{rank}(C_1 \phi_i \psi_i B_1) \leq \mathrm{rank}(\phi_i \psi_i) \leq \mathrm{rank}(\phi_i) = 1 \tag{B2}$$

where rank(*) denotes the rank of the matrix.

Since $\psi_i B_1 \neq 0$ and $C_1 \phi_i \neq 0$, we have:

$$1 \leq \mathrm{rank}(C_1 \phi_i \psi_i B_1) \tag{B3}$$

Therefore, $\mathrm{rank}(C_1 \phi_i \psi_i B_1) = 1$, and $|C_1 \phi_i \psi_i B_1|$ has $m$-1 zero roots and one non-zero root $\eta$:

$$|C_1\phi_i\psi_i B_1| = \eta \cdot 0^{m-1} \tag{B4}$$

Substituting (B4) into (B1), we get:

$$|H(\lambda_i)| = \frac{0^m}{\eta \cdot 0^{m-1}} = \frac{0}{\eta} = 0 \tag{B5}$$

When the $i$-th mode is neither controllable nor observable, $\psi_i B_1 = 0$ or $C_1\phi_i = 0$ exists, hence we have:

$$\text{rank}(C_1\phi_i\psi_i B_1) \le \min\{\text{rank}(C_1\phi),\text{rank}(\psi_i B_1)\} = 0 \tag{B6}$$

Note that $\left| D_1 + \sum_{\substack{j=1 \\ j\neq i}}^{n} \frac{C_1\phi_j\psi_j B_1}{\lambda_i - \lambda_j} \right|$ is not infinite, (18) simplifies to:

$$|H(\lambda_i)| = \frac{1}{\left| D_1 + \sum_{\substack{j=1 \\ j\neq i}}^{n} \frac{C_1\phi_j\psi_j B_1}{\lambda_i - \lambda_j} \right|} \neq 0 \tag{B7}$$

In conclusion, $|H(\lambda_i)| = 0$ holds if and only if the $i$-th mode is both controllable and observable. ■